\documentclass[twocolumn,abs,prb,longbibliography]{revtex4-1}

\usepackage{graphicx}
\usepackage{dcolumn}
\usepackage{float}
\usepackage{amsmath}
\usepackage{bm}

\newcommand{\comment}[1]{}

\begin{document}


\title{Detecting superconductivity in the high pressure hydrides and metallic hydrogen from optical properties}

\author{J. P. Carbotte$^{1,2}$}
\email{carbotte@mcmaster.ca}
\author{E. J. Nicol$^{3}$}
\email{enicol@uoguelph.ca}
\author{T. Timusk$^{1,2}$}
\affiliation{$^1$Department of Physics and Astronomy, McMaster
University, Hamilton, Ontario L8S 4M1, Canada}
\affiliation{$^2$The Canadian Institute for Advanced Research, Toronto, ON M5G 1Z8, Canada}
\affiliation{$^3$Department of Physics, University of Guelph,
Guelph, Ontario N1G 2W1, Canada} 
\date{\today}

\begin{abstract}{
We present a new technique for measuring the critical temperature $T_c$ in the high pressure, high $T_c$ electron-phonon-driven superconducting hydrides. This technique does not require connecting leads to the sample. In the multiphonon region of the absorption spectrum, the reflectance mirrors the temperature variation of the superconducting order parameter. For an appropriately chosen value of photon energy of order twice the gap plus 1.5 times the maximum phonon energy, the temperature dependence of the reflectance varies much more rapidly below $T=T_c$ than above. It increases with increasing temperature in the superconducting state while it decreases in the normal state. Examining the temperature dependence of the reflectance at a fixed photon energy, there is a cusp at $T=T_c$ which provides a measurement of the critical temperature. We discuss these issues within the context of the recently observed metallic phase of hydrogen. 
}
\end{abstract}

\maketitle


Currently, the highest value of superconducting transition temperature confirmed experimentally through resistivity and Meissner techniques is $T_c=203$ K in H$_3$S under a pressure of 155 GPa.\cite{Drozdov:2015,Drozdov:2014} This large critical temperature was anticipated in density functional theory (DFT) calculations\cite{Duan:2014} of the band structure, phonon dynamics and electron-phonon spectral density, as was the crystal structure Im-3m which was confirmed in synchrotron X-ray diffraction experiments.\cite{Einaga:2016} At the same time, many theoretical papers appeared which supported the idea of conventional superconductivity.\cite{Errea:2015,Bernstein:2015,Papaconstantopoulos:2015,Flores-Livas:2016,Nicol:2015,Errea:2016} Recently, optical spectroscopy data has revealed that an unprecedented large energy scale\cite{Capitani:2017} is involved in the pairing which provides confirmation of a conventional electron-phonon mechanism but with strong coupling to the hydrogen vibrational modes extending to $\approx 250$ meV. There have been many other predictions of superconductivity in the hydrides under pressure based on DFT. An example is LaH$_{10}$ which has been synthesised in the laboratory.\cite{Liu:2017} Although its $T_c$ is expected to be of order 260 K, it has yet to be measured. Another example is H$_3$P with a measured\cite{Drozdov:2015b}
$T_c\approx 100$ K in agreement with DFT calculations.\cite{Oh:2016} Also notable is solid metallic hydrogen which has a long history including early work by Ashcroft.\cite{Ashcroft:1968} Borinaga {\it et al.}\cite{Borinaga:2016} present first principle calculations of metallic hydrogen in the range of 400-600 GPa of pressure in I4$_1$/amd symmetry with a $T_c$ of 300 K at 500 GPa. They include anharmonic effects in their lattice dynamic computations. The work was extended in a further publication\cite{Borinaga:2017} to calculations of the optical conductivity which includes interband optical transitions and considers  the UV region. Degtyarenko {\it et al.}\cite{Degtyarenko:2017} present  related results of DFT-LDA (density functional theory in the local density approximation) for a slightly different structure with FDDD symmetry and found a $T_c$ of $\sim 240$ K at 500 GPa. This makes clear that state-of-the-art electronic band structure calculations  lead one to expect superconductivity in metallic hydrogen at unprecedented high temperature. A paper by McMahon and Ceperley\cite{McMahon:2011} which considers pressure in the TPa range, estimates that $T_c\sim 750$ K is an approximate upper bound for the critical temperature of metallic hydrogen. Recently, Dias and Silvera\cite{Dias:2017} have reported what is the first production of a metallic hydrogen sample under 495 GPa. It is not known whether this sample is superconducting. Dias and Silvera\cite{Dias:2017} give optical data at frequencies above $\sim 1$ eV extending up to 3 eV and find some temperature dependence between 5 K and 80 K as well as variation as a function of photon energy. However, this is not an appropriate range of energy or temperature to probe for superconductivity or for the signature of large electron-phonon inelastic scattering. In establishing the large energy scale involved in the superconductivity of H$_3$S, Capitani {\it et al.}\cite{Capitani:2017} scanned the range from 450-600 meV in photon energies and varied temperatures around $T_c$, which for superconducting metallic hydrogen would be of order 250 K, or even higher,  depending on which of the present estimates  is preferred. In the theoretical work of Borinaga {\it et al.}\cite{Borinaga:2017}, no significant  temperature dependence  to the conductivity is found up to 80 K and its variation with photon energy from 1 to 3 eV is much smaller than observed. 

In this letter, we calculate the temperature and photon energy dependence of the superconducting state optical conductivity of metallic hydrogen with emphasis on temperatures up to $T_c$ and photon energy  in the range ($2\Delta_0+\omega_{\rm max}$,$2\Delta_0+2\omega_{\rm max}$). In this photon energy range, we find important changes  in the reflectance that relate directly to the coupling of high energy phonons ($\omega_{\rm max}$ is maximum) and also establish that the temperature variation  of the reflectance is encoded  with information on the temperature dependence of the superconducting order parameter and consequently the value of $T_c$. There is an abrupt change of the variation of reflectance with temperature at the superconducting phase transition temperature which can be used to obtain an independent measurement of $T_c$, separate from the Meissner effect or drop in resistivity to zero.

For a conventional electron-phonon superconductor, the parameters that determine the properties of the superconductor are the electron-phonon spectral density $\alpha^2F(\omega)$ and the Coulomb repulsion parameter $\mu^*$. These enter the Eliashberg equations, the solutions of which in turn provide the optical conductivity\cite{Marsiglio:2008,Carbotte:1990,Akis:1991} using a Kubo formula. As we have already described, the first of these two quantities can now be reliably computed in DFT. The size of the Coulomb pseudo potential is less critical and is typically of order 0.1 or so. In all the theoretical results presented here, we have used the $\alpha^2F(\omega)$ of metallic hydrogen at 500 GPa given in Ref.~\onlinecite{Degtyarenko:2017} and have taken their value of $T_c=240$ K as representative. Results do not depend importantly on specific details.

\begin{figure}
\includegraphics[width=0.9\linewidth]{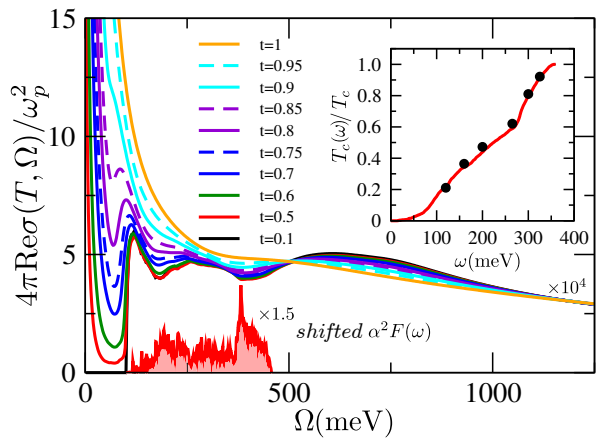}
\caption{The real part of the superconducting state optical conductivity $4\pi\sigma(T,\Omega)/\omega^2_p$ as a function of photon energy $\Omega$ for various values of temperature relative to $T_c$, with $t=T/T_c$. The conductivity is multiplied by $10^4$ for this plot and $\omega_p$ is the plasma frequency. The red distribution is the electron-phonon spectral density $\alpha^2F(\omega)$ displaced by twice the energy gap at zero temperature $2\Delta_0$ and scaled up by a factor of 1.5 on this plot. The spectral density is from Ref.~\onlinecite{Degtyarenko:2017} for metallic hydrogen at 500 GPa with $T_c=240$ K. The inset shows the value of $T_c(\omega)$ (solid black dots) obtained from our Eliashberg calculations when only the phonon modes below $\omega$ are included. The red curve is the partial area under $\alpha^2F(\omega)$ up to $\omega$ divided by the total area under $\alpha^2F(\omega)$ integrated up to $\omega_{\rm max}\sim 360$ meV.
}\label{fig1}
\end{figure}
In Fig.~\ref{fig1}, we show the absorptive part of the superconducting state optical conductivity at many values of temperature as labelled on the figure where $t=T/T_c$. Also shown is the electron-phonon spectral density $\alpha^2F(\omega)$ of Ref.~\onlinecite{Degtyarenko:2017} on which our Eliashberg calculations are based. The spectral density plotted is shifted to the right by $2\Delta_0$ (red distribution, lower part of the figure), to illustrate how dips in the low $T$ conductivity line up with major peaks in this spectrum. The calculated value of the zero temperature gap $\Delta_0=\Delta(T=0)$ is 50 meV, which gives a ratio $2\Delta_0/k_BT_c=4.85$, very close to the value of 4.9 expected for a conventional superconducting material with characteristic phonon energy $\omega_{\ln}=1400$ K and a $T_c/\omega_{\ln}$ ratio of 0.17 (see Ref.~\onlinecite{Carbotte:1990}). The solid black curve is the conductivity at reduced $t=T/T_c=0.1$. It largely sits under the red curve for $t=0.5$. There is an optical gap $\Delta_{\rm op}$ equal to twice the superconducting gap $\Delta_0$ below  which there is no significant absorption at such low temperatures. Above this gap, there is a sharp absorption edge with the magnitude of its vertical rise determined by the size of the residual scattering that we include. Here, $1/\tau_{\rm imp}$ was chosen to be 100 meV. At higher photon energies beyond $\Delta_{\rm op}$, the inelastic scattering due to the electron-phonon spectral density shows small modulations which reflect some of the details of $\alpha^2F(\omega)$ (red distribution). As temperature is increased the absorption coming from the thermally excited Bogoliubov quasiparticles appears below the gap. These grow in importance  with increasing temperature and the gap structure  gradually becomes less prominent. This region is not optimal for pressure cell experiments due to small sample size.
At $T_c$, a Drude-like curve is obtained with a phonon-assisted region at higher energies. There is a crossing at $\Omega=500$ meV after which all superconducting curves fall above its normal state (solid orange curve). This remains to a much higher frequency. It is this region that we will exploit in what follows. In the inset of Fig.~\ref{fig1}, we show the value of the critical temperature $T_c(\omega)$ obtained from our Eliashberg calculations when only the phonons up to $\omega$ in $\alpha^2F(\omega)$ are included (solid black dots). The continuous red curve is  $\int_0^\omega\alpha^2F(\omega')d\omega'/\int_0^{\omega_{\rm max}}\alpha^2F(\omega')d\omega'$, {\it i.e.} the ratio of the partial area under the spectral density integrated up to $\omega$ divided by the entire area under the spectral density, where $\omega_{\rm max}\sim 360$ meV is the maximum phonon frequency. As about a third of the $T_c$ comes from the high phonon frequencies in the spectrum, such a high energy scale should be reflected  in superconducting properties including the optical conductivity.

\begin{figure}
\includegraphics[width=0.9\linewidth]{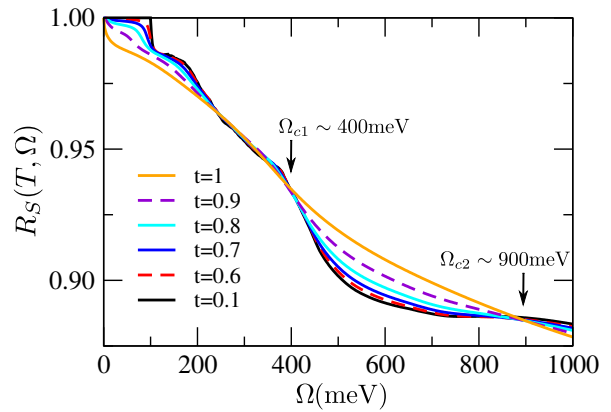}
\caption{The reflectance $R_S(T,\Omega)$ of superconducting metallic hydrogen as a function of photon energy $\Omega$ for several temperatures, with $t=T/T_c$. The arrows at $\Omega_{c1}\sim 400$ meV and $\Omega_{\rm c2}\sim 900$ meV define the range of photon energies for which the reflectance in the superconducting state increases as temperature increases. This behavior is opposite to that expected in a normal metal and is taken to be the signature of superconductivity.
}\label{fig2}
\end{figure}
In experiment, it is the reflectance $R_S(T,\Omega)$, rather than the conductivity, that is directly measured. This depends on both real (Fig.~\ref{fig1}) and imaginary part 
of the conductivity, the latter being the Kramers-Kronig transformation of the former.
$R_S(T,\Omega)$  
is presented in  Fig.~\ref{fig2} for the some of the same temperatures as Fig.~\ref{fig1}. We have taken $\omega_p=33$ eV and in calculating the reflectance we have corrected for the diamond anvil cell by using the index of refraction of $n_0=2.417$ instead of the free space value of $n_0=1$.
At low temperature $R_S(T,\Omega)$  is unity below the optical gap $\Delta_{\rm op}$. At higher energies the phonon structure highlighted in Fig.~\ref{fig2} between $\Omega_{c1}$ and $\Omega_{c2}$ is evident with the reflectance increasing with increasing temperature, a trend which is reversed above
$\Omega_{c2}$. This behaviour is a clear signature of superconductivity. We note that while there remains temperature dependence around $\Omega=1$ eV, this is very small $\sim 1$\% for a range from $T=0$ to $T=T_c$. For $T=80$ K, used in the experiments of Dias and Silvera\cite{Dias:2017}, we find no change at all from the curve at $T=5$ K. Further, the variation with $\Omega$ between $\Omega=1$ eV and 3 eV is much larger $\sim 17$\% than found in experiment\cite{Dias:2017}. Both these estimates are made without including the effect of the diamond anvil cell.

\begin{figure}
\includegraphics[width=0.9\linewidth]{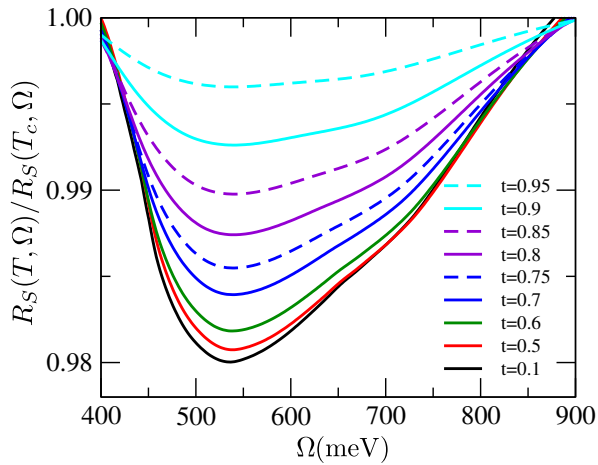}
\caption{The normalised reflectance $R_S(T,\Omega)/R_S(T_c,\Omega)$ as a function of photon energy in the interval $\Omega_{\rm c1}=400$ meV to $\Omega_{\rm c2}=900$ meV indicated by arrows in Fig.~\ref{fig2}. In this energy range, the reflectance increases with increasing temperature, where $t=T/T_c$. The increase is small at low temperature but accelerates as $T$ approaches $T_c$. This variation reflects directly the temperature dependence of the order parameter (energy gap) associated with the second order phase transition of BCS theory.
}\label{fig3}
\end{figure}
Here we are most interested in the lower energy region $\Omega<\Omega_{c2}$ not reported in Ref.~\onlinecite{Dias:2017}. It is this region which is most optimum for the detection of superconductivity using optics. In Fig.~\ref{fig3}, we show the superconducting state reflectance $R_S(T,\Omega)$ normalised to its value at $T_c$ as a function of photon energies between 400 meV and 900 meV. As was pointed out in Ref.~\onlinecite{Capitani:2017} on H$_3$S, in this region the reflectance increases with increasing temperature  opposite to what is seen in conventional metals. This unusual effect is directly traced to the electron-phonon mechanism with strong coupling to high energy phonons. The magnitude of the changes predicted for the hydrogen are of the same order as was measured in H$_3$S.
We also note that the curves vary most rapidly as temperature is increased towards $T_c$. The curve at $t=0.5$ is virtually the same as for zero temperature. This allows us to use these curves to get a measure of $T_c$ directly from optics, independent  of measurements of zero resistance  or the Meissner effect. 

\begin{figure}
\includegraphics[width=0.9\linewidth]{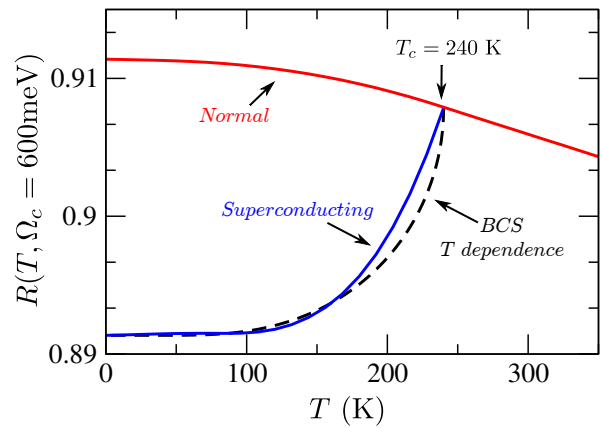}
\caption{The reflectance of metallic hydrogen at fixed photon energy $\Omega_c=600$ meV as a function of temperature. The solid red curve is the normal state while the solid blue is for the superconducting state. The temperature at which these two curves meet defines the value of the critical temperature of the sample, which here is 240 K. The dashed black curve is for comparison and is a scaling of the mean field temperature dependence of BCS theory, {\it i.e.} $[a(1-b\Delta(T)/\Delta_0)]$, where $a$ and $b$ are scaling constants. 
}\label{fig4}
\end{figure}
In Fig.~\ref{fig4}, we plot the value of the unnormalised $R(T,\Omega_c)$ where we have chosen $\Omega_c=600$ meV which is approximately $2\Delta_0+{3\over 2}\omega_{\rm max}$. This photon energy falls in the region of maximum difference in the normalised curves of Fig.~\ref{fig3}. The solid blue curve gives our result for the superconducting state $R_S$ while the red curve is for the underlying normal state $R_N$. We see the expected flat part in $R_S(T,\Omega_c)$ at small $T$ and a much more rapid rise as $T_c$ is approached. By contrast, our normal state results (red curve) vary much less with temperature and show a small decrease with increasing temperature instead of the rapid increase of the blue curve. In such a plot $T_c$ is the temperature at which the two curves meet. The value of $T_c$ need not be known from some other measurement but will correspond to a cusp in the plot of Fig.~\ref{fig4}, where below the cusp, the reflectance is rapidly increasing with $T$ while above, it is decreasing with variation on a much reduced scale. To get some understanding of this behaviour, we have plotted as the dashed black curve in Fig.~\ref{fig4}, the mean field  temperature dependence of the energy gap of BCS theory, {\it i.e.} $a[1-b\Delta(T)/\Delta_0]$, with $a$ and $b$ adjusted to give the values of $R_S(T=0,\Omega_c)$ and $R_S(T=T_c,\Omega_c)$. Comparing dashed and solid blue curves shows that $R_S(T,\Omega_c)$ knows about the closing of the superconducting gap at $T=T_c$ and the cusp at $T_c$ in our curves is a direct result of the BCS second order phase transition. The energy $\Omega_c$ falls in the multiphonon range of the absorption spectrum. These processes are only possible whenever the electronic density of states has energy dependence  and so they are greatly reduced as the superconducting gap closes and finally disappear at $T=T_c$. This extra absorption decreases rather than increases as temperature is increased and this is a clear signature of the superconducting state which here we have exploited to get a measurement of $T_c$ from optical data at a high frequency $\Omega_c=600$ meV. At photon energies well above $\Omega_{c2}$, the temperature variation of $R(T,\Omega)$ in the superconducting state is only slightly different from its normal state and decreases with increasing $T$  as in the normal state so that, effectively, one cannot use this region to get information on the superconducting state.

\begin{figure}
\includegraphics[width=0.9\linewidth]{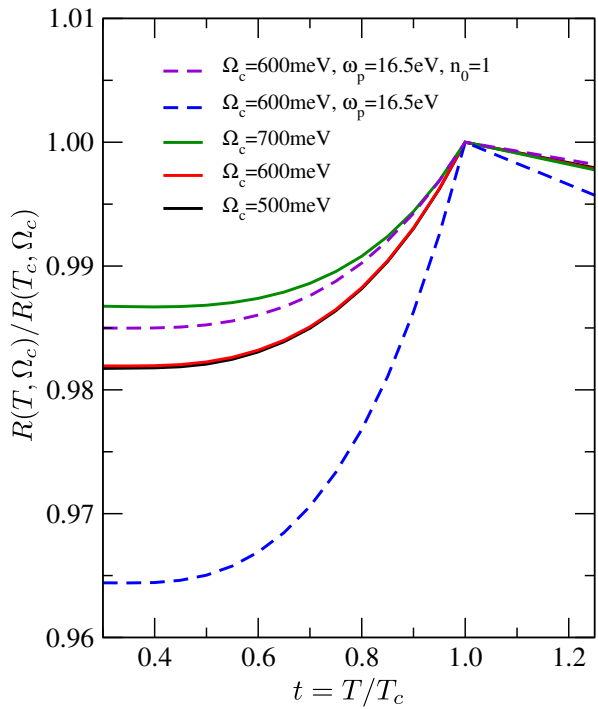}
\caption{The normalised reflectance of metallic hydrogen $R(T,\Omega_c)/R(T_c,\Omega_c)$ at fixed  photon energy $\Omega_c$ as a function of temperature. The various curves illustrate how the cusp at $T=T_c$ in this quantity changes for different choices of parameters in the reflectance calculation. The residual scattering was set at $1/\tau_{\rm imp}=100$ meV. Increasing $1/\tau_{\rm imp}$ to 500 meV produces no noticeable change in this normalised quantity. The solid curves refer to different frequencies, with $\omega_p=33$ eV and the inclusion of the diamond refractive index $n_0=2.417$ as has been used throughout this paper. The dashed curves are for fixed frequency $\Omega_c=600$ meV, but showing the effect of halving the plasma frequency and of removing the correction of the diamond $n_0$, restoring it to the free space value of $n_0=1$. 
}\label{fig5}
\end{figure}
In Fig.~\ref{fig5}, we study the robustness of the cusp at $T=T_c$ in the reflectance $R(T)$ normalised to the value at $T_c$. We show results in the range of temperature from 0.3$T_c$ to 1.25$T_c$. Several choices of photon energy $\Omega_c$ are considered, namely 500 meV (solid black curve), 600 meV (solid red curve, which is sitting almost on top of the black curve) and 700 meV (solid green). The impurity content  has been set to $1/\tau_{\rm imp}=100$ meV. Increasing $1/\tau_{\rm imp}$ to 500 meV produces no significant change. It is clear that the choice of $\Omega_c$ is also not critical provided it falls in the general  area of the broad minimum of the curves shown in Fig.~\ref{fig3}. Reducing the plasma frequency by a factor of 2 gives the dashed blue curve while, in addition, removing the correction for the diamond index of refraction $n_0=2.417$ associated with the high pressure diamond anvil cell, and replacing it with $n_0=1$, produces the dashed purple curve. In both these last cases, $\Omega_c=600$ meV. It is clear from these results  that the cusp at $T=T_c$ is robust and remains prominent in all these cases.

In summary, we have calculated the optical conductivity for superconducting metallic hydrogen within Eliashberg theory with electron-phonon parameters recently obtained from density functional theory. In the range of 1 eV to 3 eV, we find no significant variation with temperature below 80 K and a larger dependence on photon energy $\Omega$ than was found in a recent experiment. We argue that the temperature range around the critical temperature $T_c$ and photon energies around twice the gap plus 1.5, the maximum phonon energy is a much more optimum range for probing the superconducting state which results from strong coupling of the electrons to the high energy phonons. The reflectance in this range contains information on the mean field temperature  dependence of the gap which can be exploited to obtain a direct measurement of $T_c$. The reflectance, at a fixed photon frequency, has a cusp as a function of temperature which occurs at $T=T_c$ below which it increases strongly with temperature while above it decreases but on a much reduced scale. This behavior provides direct evidence of an electron-phonon mechanism.

We thank E. A. Mazur for providing the electron-phonon spectral function used in this analysis.
This work has been supported by the Natural Sciences and Engineering Council of Canada (NSERC) and by the
Canadian Institute for Advanced Research (CIFAR).


\bibliography{detecttcbib}

\end{document}